# The relation between the waveguide and overlap implementations of Kaplan's domain wall fermions

Maarten F.L. Golterman* and Yigal Shamir**

Institute for Theoretical Physics
University of California
Santa Barbara, CA 93106-4030, USA

## ABSTRACT

Recently, Narayanan and Neuberger proposed that the fermion determinant for a lattice chiral gauge theory be defined by an overlap formula. The motivation for that formula comes from Kaplan's five dimensional lattice domain wall fermions. In the case that the target continuum theory contains $4n$ chiral families, we show that the effective action defined by overlap formula is identical to the effective action of a modified waveguide model that has extra bosonic ghost fields. This raises serious questions about the viability of the overlap formula for defining chiral gauge theories on the lattice.

*permanent address: Department of Physics, Washington University, St. Louis, MO 63130, USA
**permanent address: School of Physics and Astronomy, Beverly and Raymond Sackler Faculty of Exact Sciences, Tel-Aviv University, Ramat Aviv 69978, ISRAEL

Some time ago Kaplan [1] proposed to realize four dimensional lattice chiral fermions as zero modes of a five dimensional fermion field which are bound to a four dimensional defect – the domain wall. Two specific implementations of Kaplan's idea are the waveguide model [2] and the overlap formula of Narayanan and Neuberger [3].

In the waveguide model, the four dimensional gauge field couples only to the fermions that live inside a wide band (the "waveguide") that contains the domain wall. The anti-domain wall is very far outside the boundaries of the waveguide, and the opposite-chirality fermion on the anti-domain wall does not couple to the gauge field.

In the "unitary gauge" formulation of the waveguide model, gauge invariance of the action is explicitly broken at the interface between the charged fermions that live inside the waveguide and the neutral fermions that live outside. However, since the interface is very far from the domain wall, one could hope that its effect will tend to zero in the continuum limit, provided one chooses an anomaly free fermion spectrum. Indeed, in weak coupling perturbation theory one finds that the consistent form of the anomaly is correctly reproduced for smoothly varying external fields [4, 5, 6]. This implies in particular that the continuum limit of lattice perturbation theory agrees with continuum perturbation theory for an anomaly free fermion spectrum, provided appropriate noninvariant counterterms are added.

Unfortunately, this promising picture based on a perturbative analysis is misleading. Due to the breaking of gauge invariance at the boundaries of the waveguide, the gauge degrees of freedom (*i.e.* the degrees of freedom along the gauge orbits) couple to the fermions. This interaction is not controlled by the gauge coupling, and therefore perturbation theory is not applicable. A nonperturbative study of the waveguide model [2] reveals a phase diagram very similar to the Smit-Swift model [7]. The main difference is that the symmetric strong Yukawa coupling phase of the waveguide model always contains massless fermions, whereas in the Smit-Swift model the existence of massless fermions requires fine tuning. The most important common feature is that the light spectrum is always *vectorlike*.

The waveguide model can be formulated in two mathematically equivalent ways. Apart from the "unitary gauge" formulation, there is an alternative formulation that contains a gauge group valued scalar field. In this "radially frozen Higgs" formulation the action is manifestly gauge invariant. The Higgs field is nothing but the gauge degree of freedom of the original gauge field. The Higgs formulation makes explicit the fact that the gauge degrees of freedom couple to the fermions, because the original action was not gauge invariant. The reason why perturbation theory is so misleading is that the Higgs field is wildly fluctuating. If one properly takes into account the Higgs dynamics, one finds that mirror fermions appear both inside and outside the waveguide boundary. Ultimately, the massless spectrum of both charged and neutral fermions is vectorlike if one is in a symmetric phase. In the broken phase the mirror fermions can acquire non-zero masses, but they remain in the low energy spectrum.

The study of the overlap formula, on the other hand, has so far been limited to smooth external gauge fields. Beyond that, the nonperturbative dynamics of the overlap model is poorly understood. What makes things especially complicated, is that the overlap formula cannot in general be regarded as the effective action of a local lattice theory.

The main result of this letter is the following. If the target continuum theory contains $4n$ chiral families, then the effective action defined by the overlap formula is exactly equal to the effective action of a modified waveguide model, for special values of the Higgs hopping parameter $\kappa$ and the Yukawa coupling $y$. These values are $\kappa = 0$ and $y = 1$.

Both the original and the modified waveguide models contain bosonic ghost fields. We will usually refer to the ghost fields as PV fields for short. There is, however, a qualitative difference between the ghost fields of the two models. The PV fields of the original waveguide model have a cutoff scale mass. They do not couple to the Higgs field, and so their action is gauge invariant even in the "unitary gauge" formulation. As a result, these PV field are true regulators that do not survive in the continuum limit. (To avoid confusion, let us mention here that in ref. [2] the waveguide model was studied in the "Higgs" formulation without gauge fields. In that case, the PV fields decouple entirely, and were therefore not considered).

In the modified waveguide model, on the other hand, the PV fields *do* couple to the Higgs field. The physical content of the overlap model depends on whether or not these bosonic ghosts develop a large mass. If they do, we expect the phase diagram and the massless spectrum to be qualitatively the same as in the waveguide model. This would imply that the overlap model defines a *vectorlike* theory. If the bosonic ghosts do not decouple, the situation could be even worse as the continuum limit will contain excitations that violate the spin-statistics relation. We will return to this point in our conclusions. While our results are derived for the special case of $4n$ families, we see no reason why the situation would be better when the number of families is not a multiple of four.

This letter is organized as follows. We first give a complete definition of the waveguide model. Using the "unitary gauge" formulation we calculate the effective action. Taking the limit of an infinite fifth dimension and denoting the resulting effective action of the waveguide model by $S^\infty(U)$, we show that $\exp\{-S^\infty(U)\}$ is equal to the numerator of the overlap formula. (See also ref. [6]). In the special case of $4n$ families we show that, moreover, the denominator of the overlap formula can be represented by a bosonic ghost field with a local lagrangian. In practice, this is done by replacing the original ghosts' lagrangian with a new one. We end with a summary of our results, and with a brief discussion of the light excitations of the modified waveguide model.

We would like to comment that most of the technical part of this letter is devoted to giving precise definitions of the two versions of the waveguide model, as well as the ingredients of the overlap formula. Once the precise definitions are given, one establishes the relation between different pieces of the partition function and the corresponding overlap formulae by a straightforward application of Lüscher's transfer matrix formalism [8]. The adaptation of this formalism to the case of domain walls has been discussed by Narayanan and Neuberger [3], and we will therefore skip most of the intermediate steps in the derivation.

The waveguide model is defined by the following partition function

$$Z = \mathcal{N} \int \mathcal{D}U_{x,\mu} \, \mathcal{D}V_x \, \mathcal{D}\phi_{x,s} \, \mathcal{D}\phi^{\dagger}_{x,s} \, \mathcal{D}\psi_{x,s} \, \mathcal{D}\bar{\psi}_{x,s} \, e^{-S} \, . \tag{1}$$

Notice that the gauge field $U_{x,\mu}$ and the Higgs field, denoted by $V_x$, are four dimensional. Both take values in some compact Lie group $G$ (typically SU(N) or U(1)). The fermions and the PV fields $\phi_{x,s}$ are five dimensional. $s$ is the fifth coordinate. The PV fields carry the same spinor and gauge group indices as the fermions. The notation $\int \mathcal{D}U_{x,\mu}$ ($\mu$ runs from 1 to 4) is a shorthand for $\prod_{x,\mu} \int dU_{x,\mu}$. A similar statement applies to other measure factors. For the PV fields we define $\int \mathcal{D}\phi_{x,s} \, \mathcal{D}\phi^{\dagger}_{x,s} = \prod_{x,s,i} \left( \int d\phi_{x,s,i} d\phi^{\dagger}_{x,s,i}/(2\pi) \right)$, where $i$ runs over the other indices of the PV fields. The normalization constant $\mathcal{N}$, to be defined later, is independent of the size of the fifth direction.

The action is

$$S = S_G(U) + S_H(V, U) + S_F(\bar{\psi}, \psi, U, V) + S_{PV}(\phi^{\dagger}, \phi, U) \, . \tag{2}$$

The precise form of the gauge field action is irrelevant for the following discussion. In order to expose the relation to the overlap model we will set $S_H(V, U) = 0$, i.e. no kinetic term for the $V$-field.

The fermions live on a five dimensional lattice that has $4L$ sites in the fifth direction. The coordinate $s$ takes values $s = -2L + 1, \ldots, 2L$, and we will use antiperiodic boundary conditions in this direction. The fifth direction is divided into four regions roughly as follows: the mass $m$ has the same sign as $s$, and the gauge field couples to fermions only on the half space $-L + 1 \leq s \leq L$, whose center is the domain wall at the origin. In general there may be several *irreps* of fermions (and PV-s). For a given *irrep*, the fermion action is

$$S_F = \sum_{x,y,s,s'} \overline{\psi}_{x,s} D_{x,s;y,s'} \psi_{y,s'},  \tag{3}$$

where the fermionic matrix is defined by

$$D_{x,s;y,s'} = \delta_{s,s'} D^{\|}_{x,y} + \delta_{x,y} D^{\perp}_{s,s'}.  \tag{4}$$

The four dimensional part is

$$D^{\|}_{x,y} = \begin{cases} \hat{D}^{\|}_{x,y}(I,-m), & s = -2L+1, \ldots, -L, \\ \hat{D}^{\|}_{x,y}(U,-m), & s = -L+1, \ldots, 0, \\ \hat{D}^{\|}_{x,y}(U,m), & s = 1, \ldots, L, \\ \hat{D}^{\|}_{x,y}(I,m), & s = L+1, \ldots, 2L, \end{cases}  \tag{5}$$

where

$$\hat{D}^{\|}_{x,y}(U,m) = (m-5)\delta_{x,y} + \frac{1}{2}\sum_{\mu=1}^{4}\left((1+\gamma_\mu)U_{x,\mu}\delta_{x+\hat{\mu},y} + (1-\gamma_\mu)U^{\dagger}_{y,\mu}\delta_{x-\hat{\mu},y}\right).  \tag{6}$$

We assume $0 < m < 1$, see ref. [3]. The part that contains the hopping terms in the $s$-direction is

$$D^{\perp}_{s,s'} = P_R D^R_{s,s'} + P_L D^L_{s,s'}.  \tag{7}$$

Here

$$P_{R,L} = \frac{1}{2}(1 \pm \gamma_5),  \tag{8}$$

and

$$D^R_{s,s'} = \begin{cases} yV^{\dagger}_x \delta_{-L+1,s'}, & s = -L, \\ yV_x \delta_{L+1,s'}, & s = L, \\ -\delta_{-2L+1,s'}, & s = 2L, \\ \delta_{s+1,s'}, & \text{otherwise}, \end{cases}  \tag{9}$$

$$D^L_{s,s'} = \begin{cases} yV_x \delta_{-L,s'}, & s = -L+1, \\ yV^{\dagger}_x \delta_{L,s'}, & s = L+1, \\ -\delta_{2L,s'}, & s = -2L+1, \\ \delta_{s-1,s'}, & \text{otherwise}. \end{cases}  \tag{10}$$

$y$ is an arbitrary coupling at this stage. We will keep the physical four dimensional volume finite, with antiperiodic boundary conditions in the time direction. The limit $L \to \infty$ is to be taken before the thermodynamic limit.

We now turn to the PV fields. For every fermion field, we introduce four species of PV fields that carry the same set of indices as the fermion field. Each PV field lives on a five dimensional lattice with $s = 1, \ldots, L/2$ (we assume $L$ to be even). The idea is that the action of each species of PV fields will roughly correspond to the square of the Dirac operator in one of the four regions defined in eq. (5).

Let us introduce the second order operators

$$\Omega_\pm(U) = \hat{D}^\dagger_\pm(U)\hat{D}_\pm(U), \tag{11}$$

where

$$\hat{D}_{x,s;y,s';\pm}(U) = \delta_{s,s'}\hat{D}^\parallel_{x,y}(U,\pm m) + \delta_{x,y}\hat{D}^\perp_{s,s'}, \tag{12}$$

$$\hat{D}^\perp_{s,s'} = \begin{cases} P_R\,\delta_{2,s'} - P_L\,\delta_{L/2,s'}, & s = 1, \\ P_R\,\delta_{s+1,s'} + P_L\,\delta_{s-1,s'}, & 1 < s < L/2, \\ -P_R\,\delta_{1,s'} + P_L\,\delta_{L/2-1,s'}, & s = L/2. \end{cases} \tag{13}$$

Again, we chose antiperiodic boundary conditions for reasons that will become clear. The action of the four species of PV fields is

$$\begin{aligned} S_{PV} &= \sum_{x,y,s,s',\pm} \phi^\dagger_{x,s,\pm} \Omega_{x,s;y,s';\pm}(U) \phi_{y,s',\pm} \\ &+ \sum_{x,y,s,s',\pm} \phi^{0\dagger}_{x,s,\pm} \Omega_{x,s;y,s';\pm}(I) \phi^0_{y,s',\pm}. \end{aligned} \tag{14}$$

The two charged species are denoted $\phi_\pm$ and the two neutral ones are $\phi^0_\pm$. We note that the neutral PV fields are *free* fields. The effective action of the neutral PV fields subtracts out an infrared divergent constant that would otherwise be present in the full effective action.

Having defined the waveguide model, let us now look at some of its properties. The action eq. (2) is invariant under the gauge transformation

$$\begin{aligned} U_{x,\mu} &\to \omega_x U_{x,\mu} \omega^\dagger_{x+\hat{\mu}}, \\ V_x &\to \omega_x V_x, \\ \phi_{x,s,\pm} &\to \omega_x \phi_{x,s,\pm}, \\ \phi^0_{x,s,\pm} &\to \phi^0_{x,s,\pm}, \\ \psi_{x,s} &\to \begin{cases} \omega_x \psi_{x,s}, & -L+1 \leq s \leq L, \\ \psi_{x,s}, & \text{otherwise}. \end{cases} \end{aligned} \tag{15}$$

Since the $V$ field takes values in the group, we can use the separate gauge invariance of the action and the measure to eliminate this field. We thus arrive at a new expression for the partition function

$$Z = \int \mathcal{D}U_{x,\mu}\, e^{-S_G(U)} Z(U), \tag{16}$$

$$Z(U) = \mathcal{N} \int \mathcal{D}\phi_{x,s}\, \mathcal{D}\phi^\dagger_{x,s}\, \mathcal{D}\psi_{x,s}\, \mathcal{D}\bar{\psi}_{x,s}\, e^{-S_F(V_x=1)-S_{PV}}\,. \tag{17}$$

Notice that terms in the action that did not depend on $V_x$ remain gauge invariant also after the elimination of the $V$-field. This applies in particular to the PV action.

Eqs. (16) and (17) give the "unitary gauge" form of the waveguide model. Notice that $S_F$ still depends on the Yukawa coupling $y$. To demonstrate the relation with the overlap formula we will henceforth set

$$S_F = S_F(y=1, V_x=1)\,. \tag{18}$$

The actions $S_F$ and $S_{PV}$ are quadratic in the fermion and PV fields respectively. Consequently, $Z(U)$ factorizes

$$Z(U) = \mathcal{N} Z_F(U)\, Z_{PV}(U)\,, \tag{19}$$

$$Z_{PV}(U) = \prod_{\pm} Z_{PV\pm}(U)\, Z_{PV\pm}(I)\,, \tag{20}$$

in obvious notation. Our goal now is to derive transfer matrix formulae for all these partition functions.

Let us first introduce the two-by-two matrices $B_\pm(U)$ and $C(U)$ by writing (see eq. (5))

$$\hat{D}^\parallel(U, \pm m) = \begin{pmatrix} -B_\pm(U) & C(U) \\ -C^\dagger(U) & -B_\pm(U) \end{pmatrix}\,. \tag{21}$$

We also define

$$K_\pm = \begin{pmatrix} B_\pm^{-1/2} & 0 \\ C^\dagger B^{-1/2} & B_\pm^{1/2} \end{pmatrix}\,, \tag{22}$$

$$K_\pm^0 = K_\pm|_{U=I}\,, \tag{23}$$

which is well defined because $B$ is hermitian and positive. The four hamiltonians $H_\pm$ and $H_\pm^0$ are defined by

$$e^{-H_\pm} = K_\pm K_\pm^\dagger\,, \tag{24}$$

with an analogous definition for $e^{-H_\pm^0}$. Finally, introducing creation and annihilation operators $a_x^\dagger$ and $a_x$, the second quantized transfer matrices are defined by

$$T_\pm = e^{-\hat{a}^\dagger H_\pm \hat{a}}\,, \tag{25}$$

and similarly for $T_\pm^0$.

Somewhat lengthy but straightforward manipulations now give rise to the following identity

$$\begin{aligned} Z_F(U) = \det D(U) &= \left(\det B_+ \det B_- \det B_+^0 \det B_-^0\right)^L \\ &\quad \times \operatorname{tr}(T_-^0)^L (T_-)^L (T_+)^L (T_+^0)^L\,. \end{aligned} \tag{26}$$

Apart from the relatively simple generalization of allowing the mass term to change sign, all the transfer matrix formulae derived in the context of domain wall fermions are special cases of Lüscher's construction, in the sense that Lüscher allows the gauge fields on each "time" slice $s$ to be independent variables, whereas in the domain wall case the gauge field is (almost) $s$-independent. For the reader who tries to reproduce this result, we note that Lüscher's definition of the transfer matrix is slightly different, and corresponds to replacing $KK^\dagger$ by $K^\dagger K$ in eq. (24). For more detail see refs. [3, 9].

For large $L$, the partition function is dominated by the ground state projectors

$$Z_F(U) \sim \left(\det B_+ \det B_- \det B_+^0 \det B_-^0\right)^L \left(\lambda_+ \lambda_- \lambda_+^0 \lambda_-^0\right)^L$$
$$\times \langle I-|U-\rangle \langle U-|U+\rangle \langle U+|I+\rangle \langle I+|I-\rangle . \quad (27)$$

Here $\lambda_\pm$ ($\lambda_\pm^0$) are the largest eigenvalues of $T_\pm$ ($T_\pm^0$) and $|U\pm\rangle$ ($|I\pm\rangle$) are the ground states of the corresponding second quantized hamiltonians.

We next turn to the PV fields. Consider for definiteness the charged PV species. One has

$$Z_{PV\pm}(U) = \left(\det \hat{D}_\pm(U) \det \hat{D}_\pm^\dagger(U)\right)^{-1} . \quad (28)$$

Now, $\det \hat{D}_\pm(U)$ can be represented using a fermionic path integral with anti-periodic boundary conditions in the $s$ direction. The transfer matrix expression for that path integral is

$$\det \hat{D}_\pm(U) = (\det B_\pm)^{L/2} \, \text{tr} \, (T_\pm)^{L/2} . \quad (29)$$

This equation implies in particular that $\det \hat{D}_\pm(U)$ is real. As a result

$$Z_{PV\pm}(U) = (\det B_\pm)^{-L} \left(\text{tr} \, (T_\pm)^{L/2}\right)^{-2} . \quad (30)$$

For large $L$ this becomes

$$Z_{PV\pm}(U) \sim (\lambda_\pm \det B_\pm)^{-L} . \quad (31)$$

We now see that the four species of PV fields cancel the infrared divergent factors in eq. (27). Taking the limit $L \to \infty$ we finally obtain

$$Z^\infty(U) \equiv \mathcal{N} \lim_{L\to\infty} \left(Z_F(U) \prod_\pm Z_{PV\pm}(U) Z_{PV\pm}(I)\right)$$
$$= \mathcal{N} \langle I-|U-\rangle \langle U-|U+\rangle \langle U+|I+\rangle \langle I+|I-\rangle . \quad (32)$$

The last row in eq. (32) is manifestly $L$-independent. We now define the effective action $S^\infty(U) \equiv -\log Z^\infty(U)$. For the normalization constant we take $\mathcal{N} = |\langle I+|I-\rangle|^{-2}$. Putting everything together, we obtain

$$\exp\{-S^\infty(U)\} = \frac{\langle I-|U-\rangle \langle U-|U+\rangle \langle U+|I+\rangle}{\langle I-|I+\rangle} . \quad (33)$$

The effective action $S^\infty(U)$ is finite in the thermodynamic limit.

This is the appropriate place to introduce the overlap formula. Denoting the effective action related to the overlap formula by $\bar{S}$ one has by definition [3]

$$\exp\{-\bar{S}(U)\} \equiv \frac{\langle I-|U-\rangle \langle U-|U+\rangle \langle U+|I+\rangle}{|\langle I-|U-\rangle| \langle I-|I+\rangle |\langle U+|I+\rangle|}. \tag{34}$$

Evidently, in the limit $L \to \infty$, the effective action of the waveguide model is equal to minus the logarithm of the numerator of the overlap formula, up to an irrelevant additive constant.

In general, the denominator of the overlap formula cannot be represented by a local lattice theory. Such a representation exists, however, in the special case that the target continuum theory contains $4n$ chiral families. Taking $n = 1$ for simplicity, the effective action that should give rise to this chiral gauge theory is $4\bar{S}$ according to ref. [3]. We will now show that $4\bar{S}$ is the effective action of a modified waveguide model.

The modified waveguide model contains four families of fermions. This means that in the general case of several *irreps*, there will be four five-dimensional fermion fields instead of one for each *irrep*. Denoting quantities that belong to the modified model by a tilde, the contribution of the four families of fermions is therefore simply

$$\tilde{Z}_F(U) = Z_F^4(U). \tag{35}$$

The main difference between the original and modified waveguide models is in the choice of PV fields. Instead of the original four species of gauge invariant PV fields we now have only two species of PV fields denoted $\tilde{\phi}_\pm$, whose action is not gauge invariant. (Equivalently, the new PV action *depends* on the $V$-field in the "Higgs" formulation).

Like the fermion field, the new PV fields live on a five dimensional lattice that contains a waveguide. The $s$-range is the same as for the fermions, i.e. $s = -2L + 1, \ldots, 2L$. The difference is that the mass in the PV action is constant: there is no domain wall or anti-domain wall for the PV fields. The new PV action is

$$\tilde{S}_{PV} = \sum_{x,y,s,s',\pm} \tilde{\phi}^\dagger_{x,s,\pm} \tilde{\Omega}_{x,s;y,s';\pm} \tilde{\phi}_{y,s',\pm}, \tag{36}$$

$$\tilde{\Omega}_\pm = \tilde{D}^\dagger_\pm \tilde{D}_\pm, \tag{37}$$

$$\tilde{D}_{x,s;y,s';\pm} = \delta_{s,s'} \tilde{D}^\parallel_{x,y,\pm} + \delta_{x,y} D^\perp_{s,s'}(y=1, V_x=1), \tag{38}$$

$$\tilde{D}^\parallel_{x,y,\pm} = \begin{cases} \hat{D}^\parallel_{x,y}(U, \pm m), & s = -L+1, \ldots, L, \\ \hat{D}^\parallel_{x,y}(I, \pm m), & \text{otherwise}. \end{cases} \tag{39}$$

Following the same reasoning as in the original waveguide model one finds

$$\tilde{Z}_{PV\pm}(U) = \left(\det B_\pm \det B_\pm^0\right)^{-4L} \left(\text{tr}\,(T_\pm)^{2L}(T_\pm^0)^{2L}\right)^{-2}. \qquad (40)$$

For large $L$ this becomes

$$\tilde{Z}_{PV\pm}(U) \sim \left(\lambda_\pm \lambda_\pm^0 \det B_\pm \det B_\pm^0\right)^{-4L} \left(\langle I\pm|U\pm\rangle\langle U\pm|I\pm\rangle\right)^{-2}. \qquad (41)$$

We now put together eqs. (27) and (41) and take the limit $L \to \infty$. Using the normalization constant $\tilde{\mathcal{N}} = \mathcal{N}^4$, we finally find for the effective action $\tilde{S}^\infty(U)$ of the modified waveguide model

$$\exp\{-\tilde{S}^\infty(U)\} = \left(\frac{\langle I-|U-\rangle\langle U-|U+\rangle\langle U+|I+\rangle}{|\langle I-|U-\rangle|\,\langle I-|I+\rangle\,|\langle U+|I+\rangle|}\right)^4. \qquad (42)$$

This completes the proof of the equality of the overlap formula and the modified waveguide model in the four flavour case.

What we have shown, is that for four families of chiral fermions in the target continuum theory, the overlap formula for the fermion determinant can be written as a euclidean path integral in a finite four dimensional volume, with the size of the box in the fifth dimension taken to infinity. This limit exists, and the result is rigorous. The choice of four families of fermions is clearly related to the absolute values of overlaps in the denominator of the overlap formula. A slightly more complicated construction with no essential differences can be carried out for two families.

This result shows that there is no essential difference between defining the chiral fermion determinant keeping the fifth dimension "strictly infinite" [3], and defining it in a completely finite volume. The overlap definition is identical to a modified waveguide model, and all the questions that were raised and studied in the original waveguide model [2] are relevant and important here. We believe that this observation also extends to the case of one chiral fermion.

In the original waveguide model, it was argued that the theory is always vectorlike, if the dynamics of the gauge fields is taken into account [2]. The interactions of the fermions with the $V$ field, which represents the gauge degrees of freedom, lead to the appearance of mirror fermions.

The analysis of the fermion spectrum carried out in ref. [2] applies to the modified waveguide model as well. Again, the relevant question is what the spectrum of the theory looks like when the four dimensional gauge fields $U_\mu$ are turned off in the "Higgs" formulation of the model. Exactly like in the original model, we expect that mirror fermions will appear at the waveguide boundaries, and when the gauge field $U_\mu$ is turned back on, some of them will couple to the gauge field, rendering the theory vectorlike [2]. In the overlap formula, they are repesented by $\langle U+|I+\rangle$ and/or $\langle I-|U-\rangle$.

In the case of the modified waveguide model or overlap formula, the PV fields also interact with the $V$ field. We believe that due to these interactions, zeromodes located at the waveguide boundaries will develop for the PV fields as well. These zeromodes remain massless as long as the gauge invariance is not spontaneously broken. (There are no domain wall like zeromodes, because the mass does not flip sign for these fields). This is a potential disaster, since the PV fields violate the spin-statistics theorem. This problem does not arise in the original waveguide model, because there the PV fields do not couple to the $V$ field, and have masses of the order of the cutoff.

When the gauge field is turned on, the mirror fermions and the ghosts will contribute to the dynamics of the complete theory. In the case that the target continuum theory is chiral (even in the anomaly free case) these contributions obviously do not cancel. Therefore, the modified waveguide model is not unitary in this case. It is conceivable that these unitarity violations will disappear in the quantum continuum limit of the fully gauged, modified waveguide model. At this stage, this remains a nontrivial open question.

If, on the other hand, one constructs a vectorlike theory with one lefthanded and one righthanded fermion by taking the overlap formula times its complex conjugate, there is no problem with mirror fermions or ghosts. In that case the combined effects of the mirror fermions and the ghosts cancel, because the imaginary part of the effective action is identically zero.

**Acknowledgements.** We would like to thank the participants of the 1995 ITP Workshop on Lattice Chiral Fermions for many discussions. This work was supported by the National Science Foundation under grant No. PHY89-04035. MG is supported in part by the Department of Energy under contract number #DOE-2FG02-91ER40628. YS is supported in part by the US-Israel Binational Science Foundation, and the Israel Academy of Science.

**Erratum**

Eqs. (27,31,41) are not valid if the overlaps $\langle U-|U+\rangle$ and $\langle U+|I+\rangle$ vanish, which happens if the gauge field is topologically nontrivial. If this is the case, the large $L$ behavior of the fermionic partition function $Z_F(U)$ will be dominated by contributions from some excited state(s) to eq. (26) for which the product of overlaps does not vanish. The effective action for the model defined in eqs. (35–39) is therefore not given by eq. (42) in the case of topologically nontrivial gauge fields.

However, if the gauge field has trivial topology (and here we will take that to mean that $\langle U-|U+\rangle$ and $\langle U+|I+\rangle$ do not vanish [3]), eq. (42), and therefore the main result, remain valid. This includes any smooth gauge field $U_\mu^{\text{smooth}}(x)$ with zero total topological charge (and, in particular, any perturbative gauge field), as well as "rough" gauge fields of the form $V(x)U_\mu^{\text{smooth}}(x)V^\dagger(x+\mu)$, where $V(x)$ is an arbitrary lattice gauge transformation.

The equality of the overlap formula and the modified waveguide model in the topologically trivial sector is in fact sufficient to apply the analysis of the fermion spectrum developed in ref. [2] to lattice chiral gauge theories defined from the overlap formula. In that analysis, the question asked is what happens if one restricts the dynamics of the theory to that of the fermions and the trivial orbit (*i.e.* gauge configurations of the form $V(x)V^\dagger(x+\mu)$) only. In the continuum limit, the desired result is a theory of free, undoubled chiral fermions [2]. In order to address this question for the case of the overlap, we can still use the equivalence to the modified waveguide model, and the conclusions about the fermion and PV ghost spectrum remain exactly the same. For a less technical and more self-contained summary of our conclusions, we refer to the proceedings of Lattice'95.

We believe that an equivalence between the overlap formula and the $L \to \infty$ limit of a euclidean lattice partition function exists for all lattice gauge fields and we hope to return to this more general question in the near future.

**Acknowledgement.** We would like to thank R. Narayanan for a discussion leading to the discovery of the error.